\documentclass{article}

\usepackage{graphicx}
\usepackage{booktabs} 

\usepackage{amsmath,amssymb,amsfonts}
\usepackage{amstext}
\usepackage{amsthm}
\usepackage{amsbsy}
\usepackage{bbm}
\usepackage{graphics}
\usepackage{algorithm}
\usepackage[noend]{algpseudocode}
\usepackage{cite}
\usepackage{multirow}
\usepackage{pdfsync}
\usepackage{epstopdf} 
\usepackage{color}
\usepackage{caption}
\usepackage{subcaption} 
\usepackage{enumitem}
\usepackage[noend]{algpseudocode}

\floatname{algorithm}{Procedure} 

\usepackage[labelformat=simple]{subcaption}

\usepackage{hyperref}

\DeclareMathOperator*{\argmax}{argmax}

\usepackage[accepted]{icml2018}

\icmltitlerunning{Learning To Activate Relay Nodes: Deep Reinforcement Learning Approach}

\begin{document}

\twocolumn[
\icmltitle{Learning To Activate Relay Nodes: Deep Reinforcement Learning Approach}


\icmlsetsymbol{equal}{*}

\begin{icmlauthorlist}
\icmlauthor{Minhae Kwon}{equal,baylor,rice,ewha}
\icmlauthor{Juhyeon Lee}{equal,lg}
\icmlauthor{Hyunggon Park}{ewha}
\end{icmlauthorlist}

\icmlaffiliation{baylor}{Department of Neuroscience, Baylor College of Medicine, Houston, TX, USA}
\icmlaffiliation{rice}{Department of Electrical and Computer Engineering, Rice University, Houston, TX, USA}
\icmlaffiliation{ewha}{Department of Electronic and Electrical Engineering, Ewha Womans University, Seoul, Korea}
\icmlaffiliation{lg}{LG Electronics Inc., Seoul, Republic of Korea}

\icmlcorrespondingauthor{Minhae Kwon}{minhae.kwon@ewhain.net}
\icmlcorrespondingauthor{Juhyeon Lee}{jh36.lee@lge.com}
\icmlcorrespondingauthor{Hyunggon Park}{hyunggon.park@ewha.ac.kr}

\icmlkeywords{Network Formation, Network Topology Design, Reinforcement Learning, Wireless Ad Hoc Networks, Mobile Relay Networks}

\vskip 0.3in
]


\printAffiliationsAndNotice{\icmlEqualContribution} 

\begin{abstract}

In this paper, we propose a distributed solution to design a multi-hop ad hoc network
where mobile relay nodes  strategically determine their wireless transmission ranges
based on a  deep reinforcement learning approach. 
We consider scenarios where only a limited
networking infrastructure is available but a large number of wireless mobile relay
nodes are deployed in building a multi-hop ad hoc network to deliver source data to
the destination. 
A mobile relay node is considered as a decision-making agent that strategically
determines its transmission range 
in a way that maximizes network throughput while  
minimizing the corresponding transmission power consumption.
Each relay node collects information from its partial
observations and 
learns its environment through a sequence of experiences.
Hence, the proposed solution requires only a minimal amount of information
from the system.  
We show that the actions that the relay nodes take from its policy are determined 
as to activate or inactivate its
transmission, i.e., 
only necessary relay nodes are
activated with the maximum transmit power, and   
nonessential nodes are deactivated to minimize power consumption. 
Using extensive experiments, we confirm that the proposed solution builds
a network with higher 
 network performance than current  state-of-the-art solutions in terms of system goodput and connectivity ratio.

\end{abstract}

\section{Introduction}
\label{sec:intro}

\begin{figure}[tb]
\centering
\includegraphics[width = 8cm]{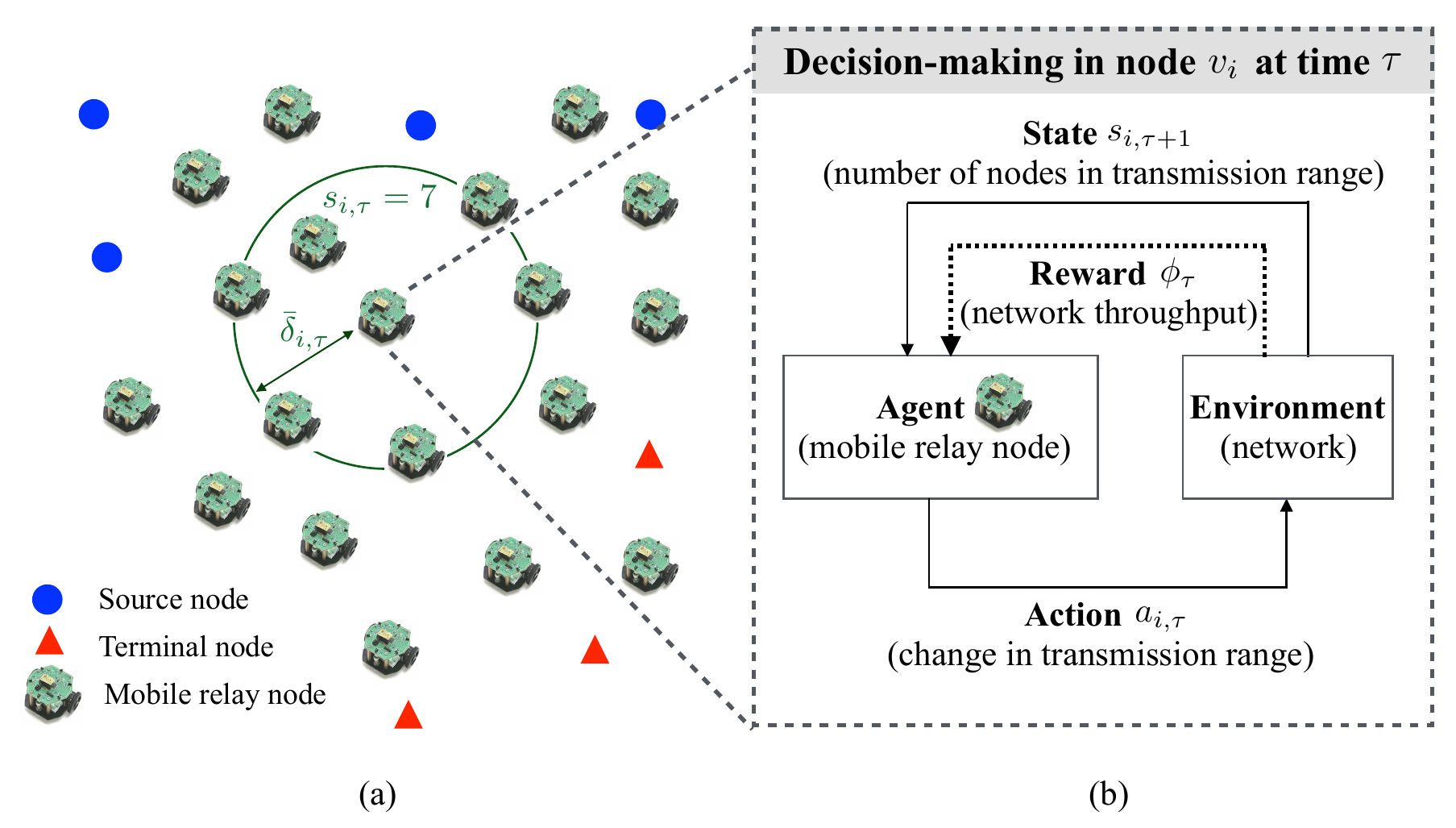}
\caption{ Illustrative examples of (a) a multi-hop wireless network, and (b) the proposed decision-making process within a node. }
\label{fig:robot}
\end{figure}

Given the rapid growth in mobile robotics, such as unmanned ground vehicles (UGVs) and
unmanned aerial vehicles (UAVs), research on autonomous network formation for
networked agent systems has gained much attention \cite{howard2002mobile,
zavlanos2008distributed, luo2012mobile, nazarzehi2018distributed}. In the perspective
of network design, the main goal is  building an energy-efficient  multi-hop network
that can connect source nodes to terminal nodes via mobile relay nodes with 
energy constraints. Thus, the network topology is determined based
on wireless connections between relay nodes. 
Possible applications include disaster networks~\cite{yuan2016robust,
erdelj2017wireless} and military networks~\cite{van2005automated} where only limited
networking infrastructure is available, and  mobile agents (e.g., robots) 
that are
sparsely spread across a large area must act as relay nodes  to deliver data in an ad
hoc manner. In such a scenario, the mobility of the agents cannot be controlled 
centrally because of  limitations in the networking infrastructure, which means the
network cannot be designed by determining the location of the mobile relay nodes.
Instead, the network must be designed by determining the wireless connections between
mobile relay nodes  roaming within preassigned areas. 
An  example of such a network is shown in Figure~\ref{fig:robot}(a).  

However,  designing an 
optimal network 
for relay nodes with mobility is challenging to do centrally because  information about the 
real-time locations of the mobile nodes 
must be collected and considered for the optimal network topology. 
The problem becomes even more challenging when  
a large number of nodes need to be 
considered.
Because the number of potential network
topologies increases exponentially with the number of nodes, the computational
complexity of finding the optimal topology becomes too high for a large-scale network to
be practically deployed. 
Therefore, in this paper, we propose a distributed solution that is practically deployable
because the mobile relay nodes 
make decisions about 
their own transmission ranges. 
Thus, the network topology is determined by the decisions that  
relay nodes make about themselves. 

Unlike a centralized solution, where all necessary information can be made available to 
decision-making agents, 
each relay node in the proposed solution  makes decisions  based on only partial observations 
about the overall network.  
Thus, each relay node should be able to learn about the network from its own
observations, with a minimum 
amount of information provided by the system. 
We  adopt 
the deep
reinforcement
learning approach~\cite{mnih2015human, silver2016mastering, van2016deep} for 
sequential decisions, allowing each 
relay node to learn its environment (i.e., network) from a sequence of experiences. 
In the proposed decision-making framework, each relay node chooses the  action
that maximizes the estimated cumulative future rewards at its state. 
Specifically, each relay node makes a decision (i.e.,  action) of 
how much   
its transmission range needs to be 
increased or decreased
based on its observations of the number of nodes in
its current transmission range (i.e., state). The reward for the action is designed to
consider both throughput improvement and the amount of 
additional transmission power (See Figure ~\ref{fig:robot}(b)). 
Note that the cumulative future reward should be available for all state-action
pairs  to find the optimal policy that returns the optimal action at each 
state.
In a large-scale network with many nodes, however, the state space is
too large 
to explore all state-action pairs and  learn the 
rewards.  
Therefore, in this paper, we use a deep neural network with an input of state-action
pairs and an output of estimated cumulative future reward and train it through a
sequence of experiences.

Using an extensive set of experiments, we confirm that the proposed solution 
makes each relay
node take actions to set its transmission range as either zero or maximum. 
Therefore, this can be viewed as the \emph{activeness} of each node, i.e., 
an \emph{active} node makes a connection with the maximum transmission range and an 
\emph{inactive}
node does not make a connection. 
Thus our autonomous activation system 
automatically selects the relay nodes needed 
to deliver data from the source to the destination and 
deactivates the  transmission mode of unnecessary relay nodes to minimize
 power consumption. 

The main contributions of this paper are summarized as follows. 
\begin{itemize}
\item We propose an autonomous network formation solution that can build a multi-hop
ad hoc network in a distributed manner,   
\item We formulate a decision-making process based on the Markov Decision Process (MDP) framework such that 
 each relay node can  
determine its optimal wireless connections while  
explicitly considering the trade-off 
  between overall
  network throughput and individual transmission power consumption,
\item We propose a learning process for the proposed solution such that 
  individual relay nodes can make decisions using  
 only minimal  amount of information from the system, 
\item We adopt a deep neural network to efficiently predict the 
  cumulative future reward
  for large-scale state-actions pairs and learn the optimized policy by maximizing
  the estimated cumulative future reward, and
\item The proposed solution can activate only the few essential nodes 
   needed to build a 
  source-to-destination connection, 
leading to an  
  autonomous activation system. 
\end{itemize}

The rest of our paper is organized as follows. In Section~\ref{sec:related_works}, we briefly review related works.  We introduce a wireless network model in Section~\ref{sec:system_setup}. 
Section~\ref{sec:prop}  
provides the proposed distributed decision-making process and corresponding
procedures for the system and nodes. An extensive set of experimental results is
provided in 
Section~\ref{sec:experiment}. 
Finally, we draw conclusions in Section~\ref{sec:conclusion}.

\section{Related Works} 
\label{sec:related_works}

Network formation strategy in wireless ad hoc networks has been 
studied in the context of self-organizing
networks~\cite{sohrabi2000,kim2006}. 
A protocol design 
for the self-organization of wireless sensor networks with a large number of
static and highly energy constrained nodes is proposed 
in~\cite{sohrabi2000}. 
In~\cite{kim2006}, a self-organizing routing protocol for mobile sensor nodes is described. 
The proposed protocol declares  membership in
 clusters  as  sensors move and confirms whether a mobile sensor node can communicate
with a specific cluster head within a specific time slot. 
Even though those self-organizing protocols can provide solutions to network design, 
they require  centralized planning, which necessitates a large amount of system overhead. 

To overcome that limitation,  
distributed approaches in which  
network nodes can make their own decisions have been proposed. 
One  widely adopted distributed decision-making approach is using game theory to consider 
how individual players choose their own actions when interacting with other players. 
In~\cite{komali2008, mhkwon2017SPL},  game-theoretic, distributed topology control for wireless
transmission power is proposed for sensor networks. The purpose
of topology control is to assign a per-node optimal transmission power such that the
resulting topology can guarantee  target network connectivity. The similar 
topology control game in \cite{eidenbenz2006} aims to choose the 
optimal power level for
network nodes in an ad hoc network to ensure the desired connectivity properties.
The dynamic topology control scheme presented in~\cite{xu2016} prolongs the lifetime of a
wireless sensor network  based on a non-cooperative game. 
In~\cite{mhkwon_ICC17}, a link formation game in a network coding-deployed network is
proposed with low computational complexity to be used  in a
large-scale network. 

Even though game-theoretic approaches can provide distributed solutions with 
theoretical analysis, 
the solutions often provide only mediocre performance in practical systems because they rely on strong assumptions of perfect
information about
other players (e.g., actions, payoff functions, strategies, etc.). 
However, information about the nodes in wireless mobile networks is often private and therefore  not completely available to other nodes. Even when information is
available, it is still difficult for all nodes to have perfect information in  real-time. 
Therefore, each node should be able to \emph{learn} its environment from 
partially available information of the network.

A node's partial observations of its environment can be well-captured by the agent-environment interaction in the MDP framework and reinforcement learning~\cite{le2018deep}. 
For example,  
an MDP-based network formation strategy is proposed in~\cite{kwon2017MDP} and shows that an intermediate node can satisfy the Markov 
property by deploying network coding. Then,  policy-compliant
intermediate nodes can
find an optimal policy based on the value iteration, thereby
adaptively evolving the network topology against network
dynamics in a distributed manner.
In~\cite{hu2010qelar, bhorkar2012adaptive}, Q-learning based distributed solutions for relay nodes are proposed. By
updating the Q-values for state-action pairs, each node can learn how to act
optimally by experiencing the consequences of its actions. 
For example, a  Q-learning-based adaptive network formation protocol for energy-efficient underwater sensor networks is proposed in \cite{hu2010qelar}, and an opportunistic node-connecting strategy based on Q-learning in wireless ad hoc networks is discussed in  \cite{bhorkar2012adaptive}. 
A deep reinforcement learning-based solution is provided in \cite{valadarsky2017}; it   proposes  a network connection strategy that lowers the congestion ratio based on 
deep reinforcement learning in wired networks,
leading to a
data-driven network formation scheme. 
An experimental study of a reinforcement learning-based design for multi-hop networks is described in \cite{syed2016route}, and more machine learning-based approaches are presented in the survey paper in \cite{boutaba2018comprehensive}. 

%
%
%

Although those studies of machine learning techniques have broadened the research
agenda, some still resort to impractical scenarios requiring infeasible information (e.g., 
the state transition probability of each agent), and others do not consider the resource-limited
characteristics of the agents  in wireless mobile networks  (e.g., battery powered device).
Therefore, in this paper, we propose an energy-efficient solution based on a deep Q-network that does not require knowledge of  the state transition probability.

\section{Autonomous Nodes with Adjustable Wireless Transmission Range }
\label{sec:system_setup}

\label{sec:system}

We model a wireless mobile ad hoc network as a directed graph $\mathcal
G_\tau$ 
that consists of a set of nodes $\mathcal V(\mathcal G_\tau)$ 
and a set of directed links $\mathcal
E (\mathcal G_\tau)$ at time step  $\tau$. 
There are 
three types of nodes in the network, which are source, relay
and terminal, and 
the node  
$v_j \in \mathcal V(\mathcal G_\tau)$ can be 
one of those types.  
A source node is denoted by $v_h$ where $h \in \mathbf H$ is an index set of source
nodes. Similarly, 
the index set of terminals for the source node $v_h$ is denoted as $\mathbf
T_h$. 

In this paper, we consider the most generalized network scenario, where 
multiple
source nodes simultaneously transmit data toward their own terminal nodes, and
each source node has an independent set of terminal nodes.  
Specifically, the source node $v_h$ for $h \in \mathbf H$ aims to deliver its data to
multiple terminal nodes $v_t$ for all $t \in \mathbf T_h$. 
Therefore, $\sum_{h \in \mathbf H}| \mathbf T_h|$ flows should be simultaneously considered, 
where $|
\cdot |$ denotes the size of a set. 
The total index set of all terminals in $\mathcal G_\tau$ is denoted as $\mathbf T =
\bigcup_{h \in \mathbf H} \mathbf T_h$, and the numbers of source nodes and terminal
nodes are denoted by $N_H$ and $N_T$, respectively. 


A relay node 
$v_i$ for $ i \in \mathbf V$, where 
$\mathbf V$ denotes an
index set of  $N_V$ relay nodes, 
can receive data from nodes and forward it to  other nodes. 
Thus,  relay nodes play an essential role 
whenever a source node is unable to directly transmit data to its terminal
nodes.
We consider the relay nodes to be wireless mobile devices 
that can move around within a bounded region   
with power constraints. 
Moreover, each relay node can 
adaptively decide its 
\emph{transmission range} 
by adjusting its transmission power, which
determines the range of potential delivery of its data. 
We denote the radius of the transmission range
of $v_i$ as $\bar \delta_{i,\tau}$,   and 
the Euclidean distance between node $v_i$ and node $v_j$
at time step $\tau$ is $\delta_{i, \tau}(v_j)$.  
Then, $v_j$ is
located in the transmission range of $v_i$
if  $\delta_{i, \tau}(v_j) \le \bar \delta_{i,\tau}$, and $v_j$ can receive  data
from $v_i$. 
The radius of the transmission range of $v_i$ is bounded by its largest 
transmission range $\Delta$ as  determined by the energy constraint, i.e., 
$0\le \bar \delta_{i,\tau} \le \Delta$.

The deployment of relay nodes naturally leads to a multi-hop ad hoc network,
and thus the network performance (e.g., throughput, connectivity and energy
consumption)  depends highly on the data path in delivery. 
Because each relay node is an autonomous agent that can independently and
strategically decide its transmission range, 
the data path in delivery is determined by 
the decisions of the nodes.
In the next section, we propose a deep reinforcement learning-based decision-making process  
that allows each node to make the optimal decision.


\section{Distributed Decision-making for Autonomous Nodes}

\label{sec:prop}
In this section, we propose a sequential decision-making solution that determines the 
transmission range of relay nodes. 

\subsection{Proposed Solution for the Decision-making Process }

Each relay node strategically and repeatedly decides its transmission range by
considering the reward  for each selected action, where the reward represents 
the improvement 
of network throughput at the cost of the power required to take the action. 
The proposed decision-making model is formulated by an MDP, expressed as 
a tuple $\left< \mathbf S, \mathbf A,  R,  \gamma  \right>$, where
$\mathbf S$ is a finite state space, $\mathbf
A$ is a finite action space,  
$R \in \mathbb R $ is a reward, and $\gamma \in [0,1)$ is a discount factor.

A state $s_{i,\tau} \in \mathbf S$ of node $v_i$ represents the number of nodes
located in its transmission range at time step $\tau$, which implies the number of
nodes that can relay data from other nodes. 
Because each node is always  located within its own transmission range, 
the size of a state 
is at least $1$
and 
at most $N_V+N_T$, i.e., $1 \le s_{i,\tau} \le  N_V+N_T$. 

An action $a_{i,\tau} \in \mathbf A$ of node $v_i$ represents the difference between the radius of the transmission range at time step $\tau$ and $\tau-1$, i.e., 
\begin{equation}
a_{i,\tau} = \bar \delta_{i,\tau}  -\bar \delta_{i,\tau-1}. 
\end{equation}
If $a_{i,\tau} > 0$ (i.e., $\bar \delta_{i,\tau} > \bar \delta_{i,\tau-1}$), 
the node increases the 
transmission range.
Similarly, if $a_{i,\tau} < 0$, the node decreases the 
transmission range (i.e., $\bar \delta_{i,\tau} < \bar \delta_{i,\tau-1}$). 
The node can maintain the same transmission range by taking
action $a_{i,\tau}=0$. 
If the radius of the current transmission range is zero (i.e., $\bar \delta_{i,\tau}=0$),
the negative action (i.e., $a_{i,\tau} < 0$) does not change $\bar \delta_{i,\tau}$.
Similarly, if  a node is already set at its maximum transmission range (i.e., $\bar
\delta_{i,\tau}=\Delta$), the positive action (i.e., $a_{i,\tau} > 0$) does not
change   $\bar \delta_{i,\tau}$.

The reward $R_{i, \tau}$ of node $v_i$ at time step $\tau$ is defined 
as a quasi-linear function that 
consists of the throughput
improvement and the amount of transmission power consumption additionally required,
expressed as 
\begin{align}
R_{i, \tau} = u+\omega \cdot \eta \left( \phi_{\tau-1} - \phi_{\tau-2} \right) - (1-\omega) \cdot  a_{i,\tau-1}. 
\label{eqn:reward}
\end{align}  
In \eqref{eqn:reward}, $u$ is a constant that guarantees the reward 
to be non-negative, and
$\phi_\tau$ denotes the network throughput of $\mathcal G_{\tau}$. 
Thus, 
$\phi_{\tau-1} - \phi_{\tau-2}$ is the throughput improvement at the cost of taking
action $a_{i,\tau-1}$. The cost for the action intrinsically includes the amount of
additional transmit power consumption at the node, as
well as the penalty for causing additional inter-node interference in the network. 
Note that the  throughput improvement (i.e., $\phi_{\tau-1} - \phi_{\tau-2}$) is a
network-dependent variable  that  is identical for all relay nodes included in
the same network. On the other hand, the cost for additional power consumption (i.e.,
$a_{i, \tau-1}$) is node-dependent.    
The weight $\omega$ $(0\le \omega \le 1)$ can be used to balance 
the throughput improvement and additional power consumption.  
For example, if the goal is only to improve the throughput 
without the consideration of power consumption associated with taking action,  
$\omega = 1$ can be set. 
On the other hand, if 
$\omega = 0$, the throughput improvement is not taken into account in the reward, i.e., the  network throughput is not interesting to the node.
The scaler $\eta$ can balance the range of $ \phi_{\tau-1} - \phi_{\tau-2}$ and $a_{i,\tau-1}$.  

As a solution to the proposed MDP problem, we adopt  deep reinforcement learning.
Each node interacts with the system through a sequence of state observations, actions, and rewards. 
The goal of the node is to select actions in a fashion that maximizes its cumulative future reward. 
To estimate that cumulative future reward, we define the Q-function of node $v_i$ as, 
\begin{equation}
Q_i^\pi(s, a) = \mathbb{E}_\pi \left[ \sum_{\tau=t}^T \gamma^{\tau} R_{i, \tau} | s_{i,t} = s, a_{i,t} = a \right] 
\label{eqn:q-func}
\end{equation}
which is the expected cumulative sum of reward $R_{i, \tau}$ discounted by $\gamma$ at each time step $\tau$, achievable by a behavior policy $\pi$ after making an observation $s$ and taking an action $a$ at the time step $t$.  
$\mathbb E_\pi$ denotes the expected value when the policy $\pi$ is used, where  the policy $\pi$ maps a state $s \in \mathbf S$ of the node 
to an action $a \in \mathbf A$ such that $a = \pi(s)$. $T$ is the horizon
of a finite MDP.


Because we consider a large-scale network that includes many wireless nodes and  induces a large state space in the MDP framework, we use a Double Deep Q-Network (DDQN; \cite{van2016deep}) to optimize the Q-function. 
The DDQN includes an online network and a target network. The parameters of the online network in the DDQN of  node $v_i$ at time step $\tau$, denoted as $\theta_{i, \tau}$, are trained to minimize the loss function  defined in \eqref{eqn:loss_func_DDQN}.  
\begin{figure*}[bt]
\begin{equation}
\mathbf \mathtt{L}(\theta_{i, \tau}) =  \left[ R_{i,\tau+1} + \gamma Q_i^\pi \left( s_{i,\tau+1}, \argmax_{a'}Q_i^{\pi} \left(s_{i, \tau+1}, a'; \theta_{i, \tau} \right); \theta_i^-\right) 
 - Q^\pi \left( s_{i,\tau}, a_{i,\tau}; \theta_{i, \tau} \right) \right]^2
\label{eqn:loss_func_DDQN}
\end{equation}
\hrule
\end{figure*}
Here, $\theta_{i}^-$ represents the parameter of the target network, which is updated every $\Xi$ steps, where $\Xi$ is the update interval for the target network. 
The DDQN evaluates the policy according to the online network, and uses the target network to compute the target Q-value.

The optimal policy $\pi^*$ is defined to maximize 
the Q-function and it returns the optimal action $a^*$ for a given state $s$, i.e., 
\begin{equation}
a^* = \pi^*(s) = \argmax_a Q_i^{\pi^*}(s, a; \theta_{i, \tau}). 
\end{equation}
In this system,  the $\epsilon$-greedy policy is used such that 
the node chooses the action that follows the obtained policy with a probability of $ 1-\epsilon $, and it chooses an action uniformly at random, otherwise. 

In the next section, we describe procedures how the proposed 
decision-making solution can be adopted 
in our autonomous node activation system. 

\subsection{Procedures of the Proposed System }

The proposed system includes an online learning algorithm for the relay nodes so that they
learn their own policy in  real time. 

Each node owns a DDQN to build its own policy. 
Specifically, 
all nodes determine their transmission range at every time step, and this
forms the network topology $\mathcal G_\tau$ at time $\tau$. Then, the network
throughput $\phi_\tau$ measured from the network topology $\mathcal G_\tau$ is 
broadcast to all relay
nodes. 
Except for the network throughput $\phi_\tau$, which is  
shared with all nodes, 
all the other parameters used in procedures such as $s_{i, \tau}$, $a_{i, \tau}$, $\bar\delta_{i,
\tau}$,  $\theta_i^-$, $\theta_{i, \tau}$  are local, i.e., 
they are used only at node $v_i$ and not shared with
other nodes. 
In this way,  
the role of the system controller is minimized in the proposed system;  all the individual nodes  participate actively in the proposed system, 
which leads to a decentralized solution.

\begin{figure}[tb]
\centering
\includegraphics[width = 8cm]{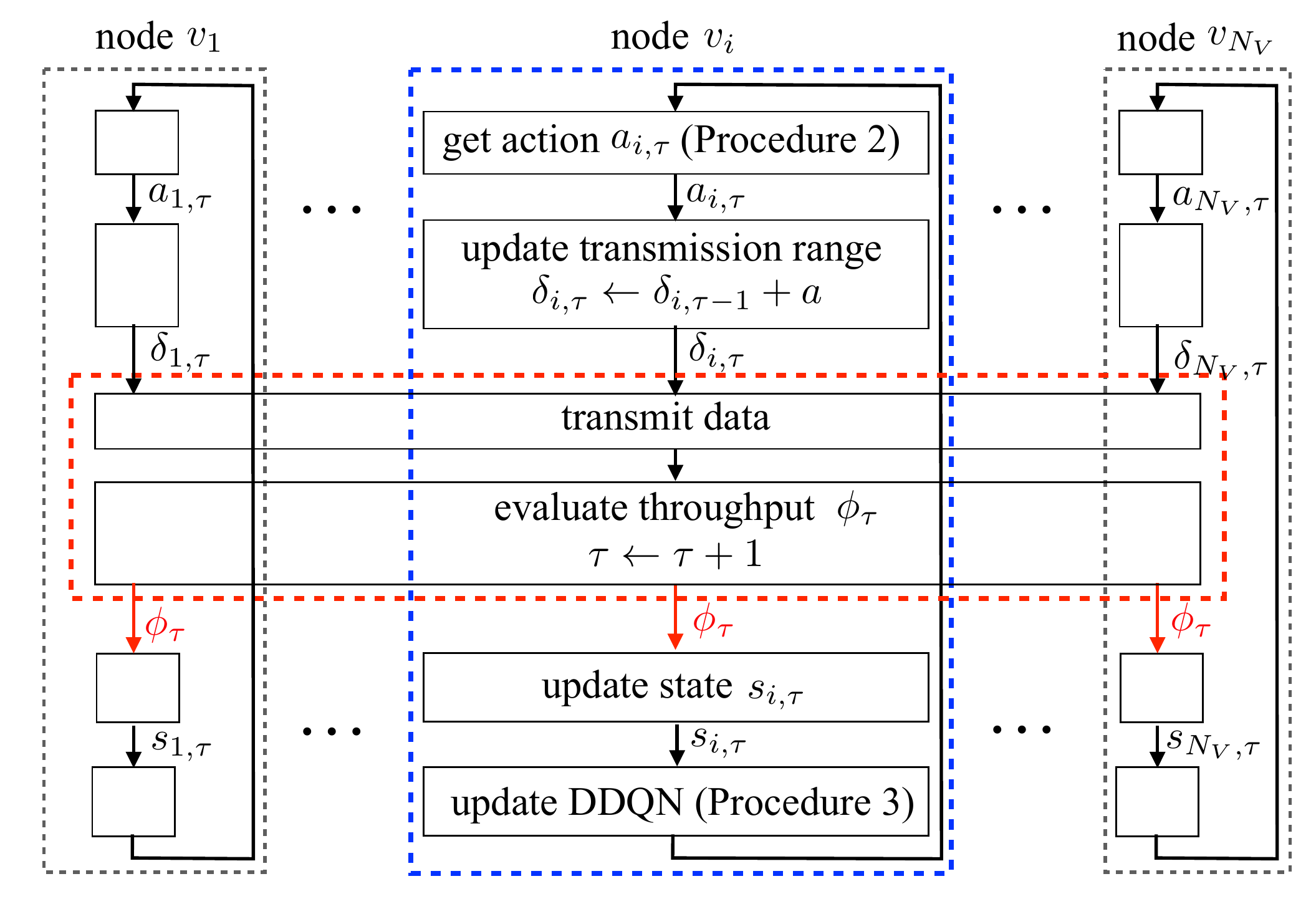}
\caption{The overall flow of the procedures performed by individual nodes in the proposed
system. 
The function blocks denoted  
by the blue dotted line  are operated by node $v_i$ 
and those in red are operated by the system.}
\label{fig:system_overview}
\end{figure}

The overall flow of the procedures performed by individual nodes in the proposed system 
is shown in Figure~\ref{fig:system_overview}. The detailed description of the 
procedures is provided in Procedure~\ref{alg:network} in conjunction with 
the subfunctions described in Procedure~\ref{alg:get_action} 
and Procedure~\ref{alg:update_dqn}.




%

\begin{algorithm}[t]
   \caption{Autonomous node activation system }
   \label{alg:network}
\begin{algorithmic}[1]
\Require{node set $\mathcal V(\mathcal G_\tau)$, action set $\mathbf A$, greedy rate $\epsilon$, target network update interval $\Xi$}
\State \textbf{Initialization:} $\phi_0 =0$, $\tau =1$ 
\State \hspace{1em}\textbf{for}{ $\forall v_i \in \mathcal V(\mathcal G_1)$} \textbf{do}
\State \hspace{2em} Initialize $\theta_{i,1}$, $\theta_i^-$ 
\State \hspace{2em} $\bar\delta_{i, 0}=0$, $s_{i,1}=0$ 
 \Repeat
\For {$\forall v_i \in \mathcal V(\mathcal G_\tau)$}
   \State $a_{i,\tau}\leftarrow \text{Get$\_$Action}$($s_{i,\tau}, \theta_{i,\tau}, \epsilon, \mathbf A$)
   \Statex\Comment{Procedure~\ref{alg:get_action}} 
   \State $\bar\delta_{i, \tau} \leftarrow \bar\delta_{i, \tau-1} + a_{i,\tau}$  
\EndFor
\State Transmits data 
\State Evaluate $\phi_\tau$ 
\State $\tau \leftarrow \tau+1$
\For {$\forall v_i \in \mathcal V(\mathcal G_\tau)$}
\State Update  $s_{i, \tau}$ 
\State Update$\_$DDQN($s_{i, \tau-1}$, $a_{i, \tau-1}$, $\phi_{\tau-1}$, $\phi_{\tau-2}$, $s_{i, \tau}$,   $\theta_i^-$, $\Xi$) \Comment{Procedure~\ref{alg:update_dqn}}
\EndFor
\Until{  Delivery terminated} 
\end{algorithmic}
\end{algorithm}

Procedure~\ref{alg:network}  describes 
how individual nodes determine their actions and update their policies  at every time step. 
In the initialization stage, 
each node sets up its initial transmission range and  DDQN. 
Then, each node chooses an action determined by the function described in 
Procedure~\ref{alg:get_action} and 
adjusts its transmission range.  
All  the network nodes transmit data simultaneously, and the network throughput is
evaluated. 
Finally, 
each node updates its state and DDQN based on
Procedure~\ref{alg:update_dqn}. These procedures, i.e.,  
get the action, update
transmission range, transmit data, evaluate network throughput, and update state and
DDQN, are repeated
until the delivery is terminated. 

\begin{algorithm}[t]
  \caption{ $\epsilon$-greedy action selection from DDQN of node $v_i$ }
\label{alg:get_action}
\begin{algorithmic}[1]
\Function{Get$\_$Action} {$s_{i,\tau},  \theta_{i,\tau}, \epsilon, \mathbf A$}
\State Generate random variable $p \sim U(0,1)$ 
\If{$p < \epsilon $}
\State Randomly select $a_{i, \tau}$ in  $\mathbf A $ 
\Else
\State $a_{i, \tau} \leftarrow \argmax_{a} Q^\pi(s_{i,\tau}, a; \theta_{i,\tau})$ 
\EndIf
\State \Return{$a_{i, \tau}$}
\EndFunction
\end{algorithmic}
\end{algorithm}
Procedure~\ref{alg:get_action} describes the process for getting  actions from the DDQN. 
For an $\epsilon$-greedy policy, a number $p$ is randomly chosen from a
uniform distribution in the range of [$0,1$]. If $p < \epsilon$, the node
explores a new action 
by randomly selecting an action from the action set. Otherwise, the node exploits the
trained policy by selecting the action $a$ that maximizes $Q^\pi(s_{i,\tau}, a;
\theta_{i,\tau})$. 



\begin{algorithm}[t]
   \caption{DDQN update at node $v_i$ }
   \label{alg:update_dqn}
\begin{algorithmic}[1]
   \Function{Update$\_$DDQN}{$s_{i, \tau-1}$, $a_{i, \tau-1}$, $\phi_{\tau-1}$, $\phi_{\tau-2}$, $s_{i, \tau}$,   $\theta_i^-$, $\Xi$} 
   \State Update $R_{i, \tau} $ using \eqref{eqn:reward} 
   \State $\theta_{i,\tau} \leftarrow \theta_{i,\tau-1} - \rho  \nabla  L(\theta_{i,\tau-1})$ using \eqref{eqn:loss_func_DDQN} 
\If{ $\tau \:\: \%$ $\Xi$ $= 0$ }
\State $\theta^-_i \leftarrow \theta_{i,\tau}$
\EndIf
\EndFunction
\end{algorithmic}
\end{algorithm}
Procedure~\ref{alg:update_dqn} describes how to train the DDQN. 
The node calculates the reward $R_{i, \tau}$ based on \eqref{eqn:reward} and finds the parameter $\theta_{i, \tau}$ that minimizes the loss function $L(\theta_{i, \tau-1})$ in \eqref{eqn:loss_func_DDQN}. The parameter $\theta_{i, \tau}$ is updated using the gradient descent method with a learning rate of $\rho$.
The target network is updated  every $\Xi$ time
steps  to stabilize the training. 

In the next section, we deploy the proposed procedures in a Wi-Fi Direct network and
evaluate the performance.

\section{Experiments}
\label{sec:experiment}
In these experiments, we consider a wireless mobile ad hoc network where multiple relay
nodes bridge source nodes and  terminal nodes. 
All relay nodes are policy-compliant agents, i.e., each node simply takes the actions 
dictated by its 
policy. Hence, each node works to build an optimal policy by
training its DDQN from a sequence of experiences. 






\subsection{Experiment Setup}

\begin{table}[t]
\caption{Network  parameters}
\label{table:simulation_parameter}
\vskip 0.15in
\begin{center}
\begin{small}
\begin{sc}
\begin{tabular}{c |c}
\toprule
Parameter & Value \\
\midrule
$\eta$&$1$\\
$\alpha$&2  \\ 
Channel Bandwidth&80 $MHz$\\
TX-RX Antennas&$3 \times 3$\\ 
Modulation Type&256-QAM \\
Coding Rate&5/6\\ 
Guard Interval &400$ns$\\
PHY Data Rate&1300 $Mbps$\\ 
MAC Efficiency&70$\% $\\
Throughput&910 $Mbps$\\ 
\bottomrule
\end{tabular}
\end{sc}
\end{small}
\end{center}
\vskip -.1in
\end{table}

The considered wireless network consists of two source nodes, 
two terminals and multiple mobile
relay nodes. The mobile relay nodes are connected by 
Wi-Fi Direct with IEEE 802.11ac standard MCS-9. 
The transmit power is computed based on  
a path loss model, expressed as 
\begin{displaymath}
 P_{TX}= P_{RX} \cdot {\left(\frac{4\pi}{\lambda}\cdot d\right)}^\alpha =\eta \cdot
d^\alpha
\end{displaymath}
where $P_{TX}$, $P_{RX}$, $\lambda$, and $d$ denote  
the transmit power, receive power,  wave length, and  
distance between the transmitter and receiver, respectively.
The parameters used in this experiment are specified by the  IEEE 802.11ac standard~\cite{IEEE80211ac, cisco80211ac} and 
shown in Table~\ref{table:simulation_parameter}. 


The network size is  the area of the bounded region in which the mobile nodes are allowed to  move,  and 
the number of relay nodes
located in the bounded region is determined by  
the Poisson Point Process (PPP) with a node density of $ 8 \times 10^{-3}$
$[\textrm{nodes}/m^2]$, which reflects the characteristics of mobile networks. 
Moreover, the locations of relay nodes
are randomly distributed over the bounded region, which captures  
their mobility. 
In this paper, 
we define an {\it episode} as a set of time steps with the same
network members. 
Hence, the number of relay nodes is reset 
whenever a new episode starts, and  
the location of the nodes  changes in every time step.


\begin{figure}[tb]
\centering
\begin{center}
\includegraphics[width = 8cm]{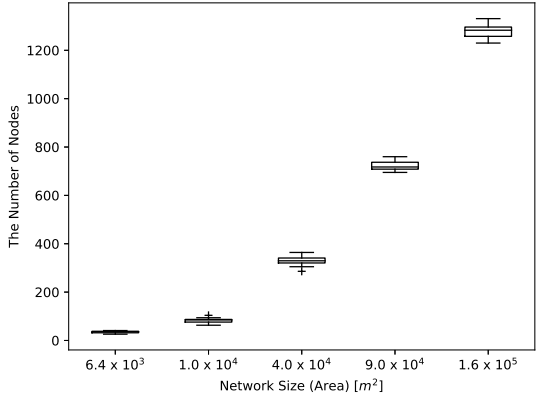}
\caption{ The number of  nodes for given network sizes with a node density of $8 \times 10^{-3}$ [nodes / m$^2$].}
\label{fig:area_numnodes}
\end{center}
\end{figure}
Figure~\ref{fig:area_numnodes} shows the number of nodes  in  $5$ networks with 
different network sizes. 
The line in the middle of  each box 
denotes the median of the experimental results, 
where on average $51$, $80$, $320$, $720$ and $1280$ nodes are located 
in networks for 
 the size of $6.4 \times 10^3 m^2$, $1.0 \times 10^4 m^2$, $4.0 \times 10^4
m^2$, $9.0 \times 10^4 m^2$, and $1.6 \times 10^5 m^2$, respectively.  The top
and bottom lines of each box represent the $25$th and $75$th
percentiles, respectively. 

The action set $\mathbf A$ has $21$ elements, with each element  
in the range of $[-1, 1]$ with a step size of 
$0.1$. The maximum radius of the transmission range $\Delta$ is set to $3.0$.
The parameters for the reward function defined 
in \eqref{eqn:reward} are set as $u = 5$, $\eta =
20$ and $\omega=0.8$. The discount factor $\gamma$ is set to  $0.7$. 
The neural network used in the DDQN has two fully connected layers, with a hidden layer size
of $32$ and ReLU nonlinearities. The neural network is optimized using the RMSProp
optimizer with a learning rate of $\rho = 0.01$. 
The target network update interval $\Xi$ is set to $100$. 
For  learning stability, we rescaled the input of the DDQN and the value of the reward by
dividing  by $100$ and $10$, respectively. 
The greedy rate $\epsilon$ decays linearly  from $1.0$ to $0.01$ in $100$ steps.


The network performance 
is evaluated based on 
the system goodput~\cite{miao2016}, defined as the
sum of data rates successfully delivered to terminal nodes, i.e., 
\begin{equation*} 
 \sum_{h = 1}^{N_H}\sum_{v_t \in \mathbf {\tilde T}_h} \frac{ L }{ \bar \tau(x_h,v_t)}
\end{equation*}
where $x_h$ denotes data generated at a source node $v_h$, 
$\mathbf {\tilde T}_h$ ($\subseteq \mathbf T_h$) denotes a set of  terminal nodes that successfully receive  $x_h$,
 $L $ represents the size of  data set $x_h$,
and $\bar \tau({x_h,v_t})$ denotes the travel time for data set $x_h$ to  arrive at 
terminal node $v_t \in \mathbf {\tilde T}_h$. 
The connectivity ratio is defined as the number of successfully connected flows out of all flows from sources to terminals.

\subsection{Experimental Results}

\begin{figure*}[tb]
    \centering
    \begin{subfigure}[b]{0.32\textwidth}
        \includegraphics[width=\textwidth]{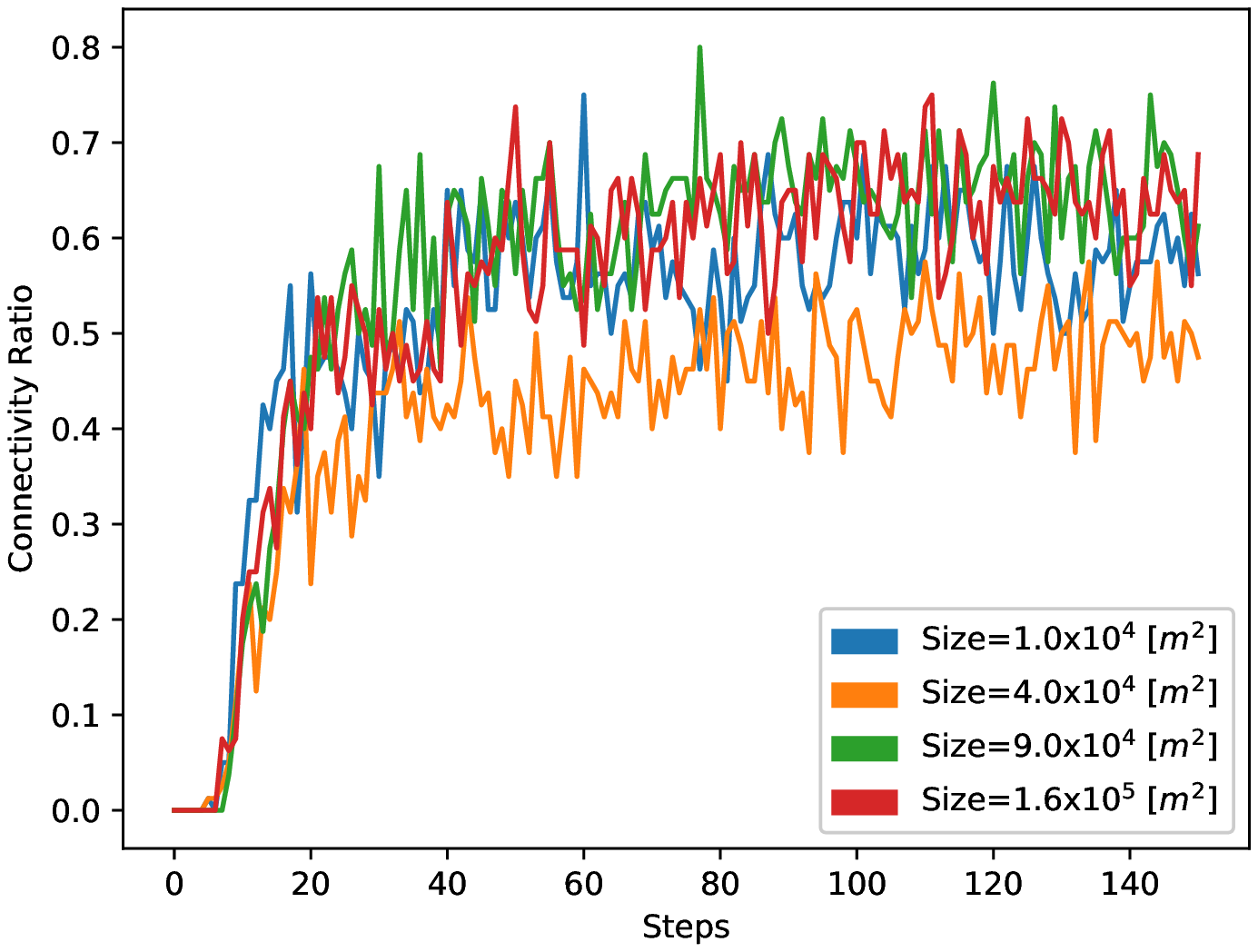}
        \caption{Connectivity Ratio}
        \label{fig:nwsize_connectivity}
    \end{subfigure}
    ~ 
    \begin{subfigure}[b]{0.32\textwidth}
        \includegraphics[width=\textwidth]{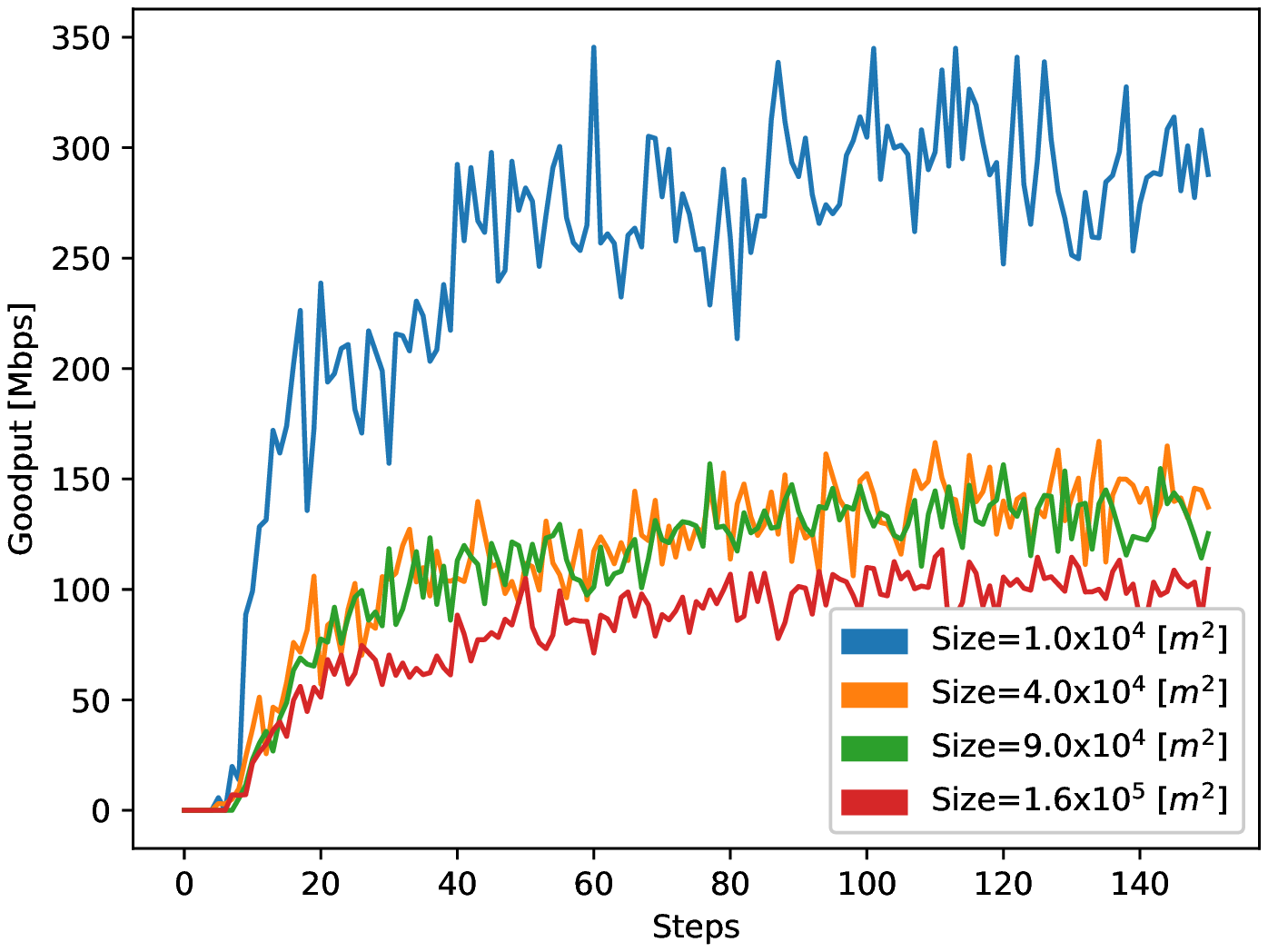}
        \caption{System Goodput [Mbps]}
        \label{fig:nwsize_goodput}
    \end{subfigure}
        ~ 
    \begin{subfigure}[b]{0.32\textwidth}
        \includegraphics[width=\textwidth]{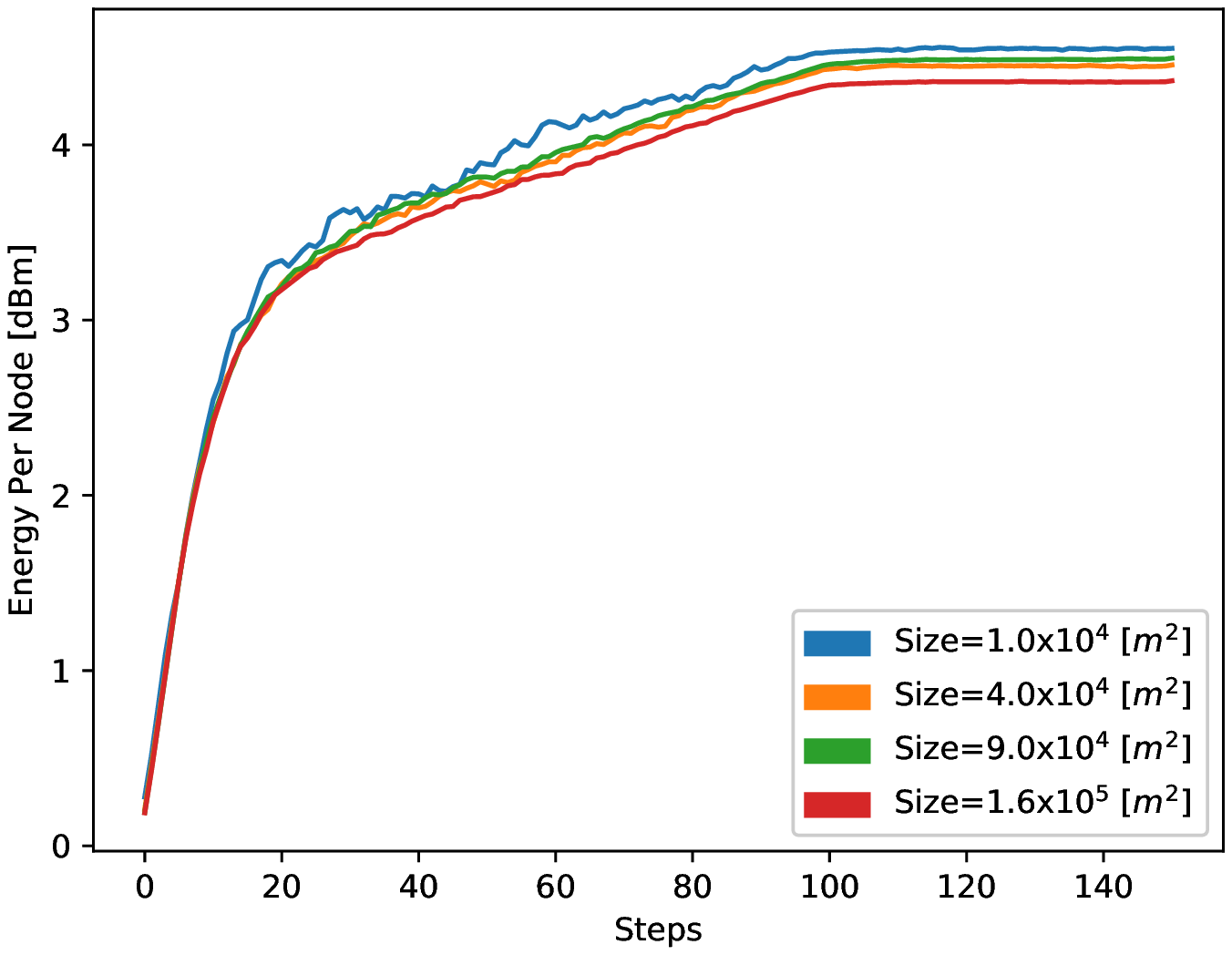}
        \caption{Power Consumption Per Node [dBm]}
        \label{fig:nwsize_energy}
    \end{subfigure}
    \caption{
    Measured performance for various network sizes     }
    \label{fig:nwsizes}
\end{figure*}

\begin{table}[tb]
\caption{Averaged results for various network sizes }
\label{table:nwsizes}
\vskip 0.15in
\begin{center}
\begin{small}
\begin{sc}
\begin{tabular}{c|ccc}
\toprule
Network & Connectivity & System  & Power \\
 Size &Ratio& Goodput&  Consumption \\
 $[m^2]$& &  [$Mbps$]& Per Node \\
  & &  &  [$dBm$] \\
\midrule
$1.0 \times 10^4$&	0.59	&291.2&	4.55\\
$4.0 \times 10^4$&	0.49	&145.6&	4.45\\
$9.0 \times 10^4$&	0.65	&136.5&	4.48\\
$1.6 \times 10^5$&	0.65	&100.1&	4.36\\
\bottomrule
\end{tabular}
\end{sc}
\end{small}
\end{center}
\vskip -.1in
\end{table}
The three network performance measures, 
 connectivity ratio, system goodput and
power
consumption per  node, are evaluated over time for networks with 
four different sizes. 
Figure~\ref{fig:nwsizes} presents the average results from $20$ independent episodes
and Table~\ref{table:nwsizes} shows the average results in the range of $[100, 150]$ time
steps (i.e., after $\epsilon$ is decayed to $0.01$).

As shown by the results  in Figure~\ref{fig:nwsize_connectivity} and 
Figure~\ref{fig:nwsize_energy},  
the network size does not significantly affect the 
connectivity ratio
or energy consumption per
node. 
Thus  the proposed system is
resilient against  network size, and can therefore be a 
scalable solution for network formation.
The advantage of scalability comes from 
the characteristics of a distributed solution in which 
individual nodes need not consider the size of the network. 
Instead, each node 
considers only its own actions and feedback from the network. Thus, the effects of 
 network size 
on the 
decision-making of each node 
are significantly limited.
On the other hand, the system goodput decreases as
the network size increases, as shown in 
Figure~\ref{fig:nwsize_goodput}, 
 because the number
of hops that the data traverse to reach a terminal increases,
as the network size increases. 

The network performance achieved by the proposed solution can be understood by
investigating how the 
transmission range of individual nodes, as well as the resulting network, are by the learning process  
over time. 
\begin{figure}[tb]
\centering
\begin{center}
\includegraphics[width = 8cm]{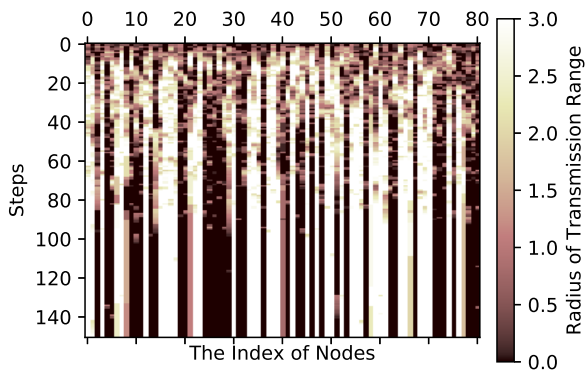}
\caption{Changes in the radius of the transmission range determined  over time by 
learning  in the network nodes }
\label{fig:color}
\end{center}
\end{figure}
In Figure~\ref{fig:color}, the radius of the transmission range is visualized for $80$
nodes over $150$ time steps. 
\begin{figure*}[tb]
    \centering
    \begin{subfigure}[b]{0.48\textwidth}
        \includegraphics[width=\textwidth]{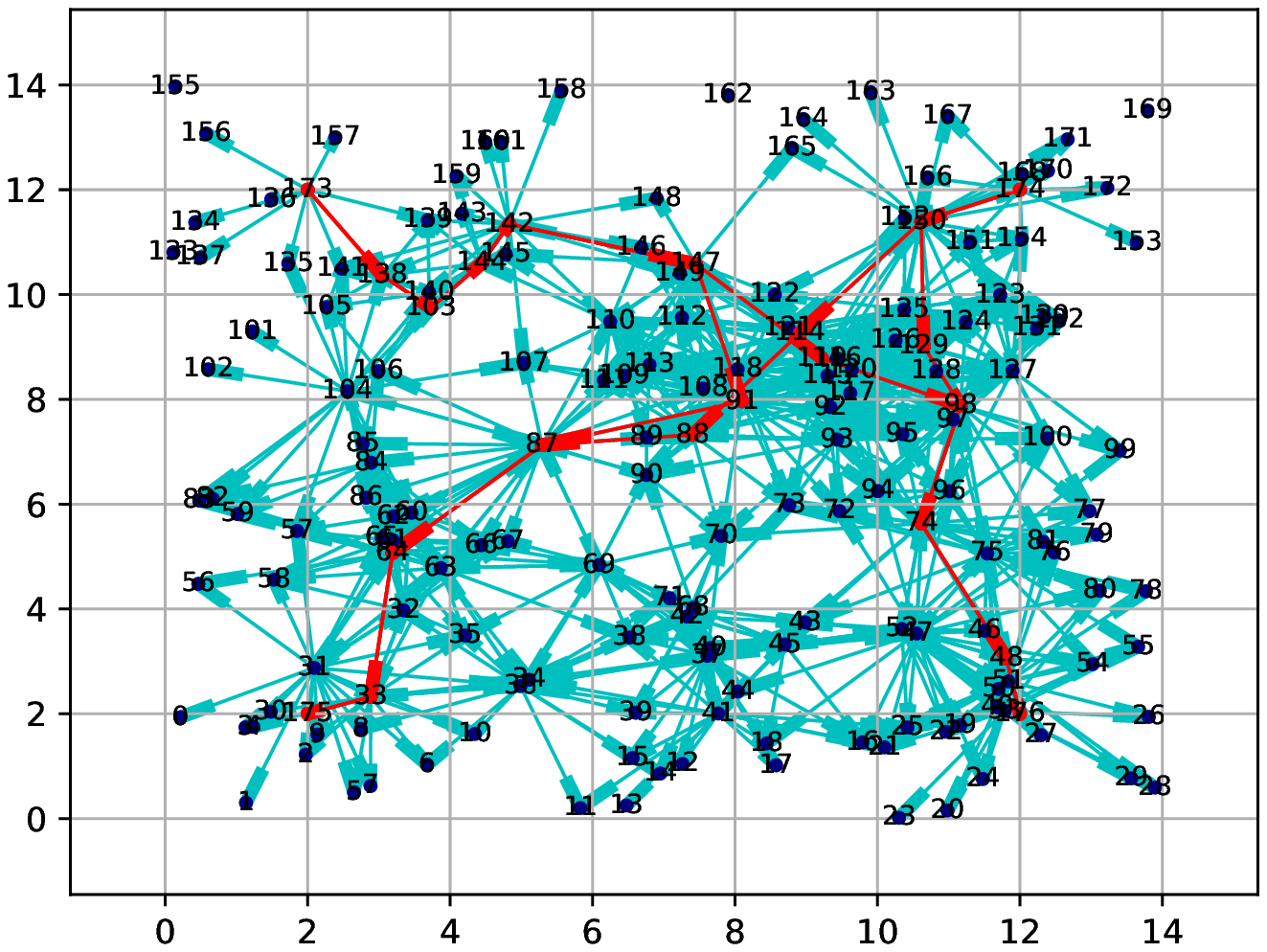}
        \caption{Step 30}
        \label{fig:step30}
    \end{subfigure}
    ~ 
    \begin{subfigure}[b]{0.48\textwidth}
        \includegraphics[width=\textwidth]{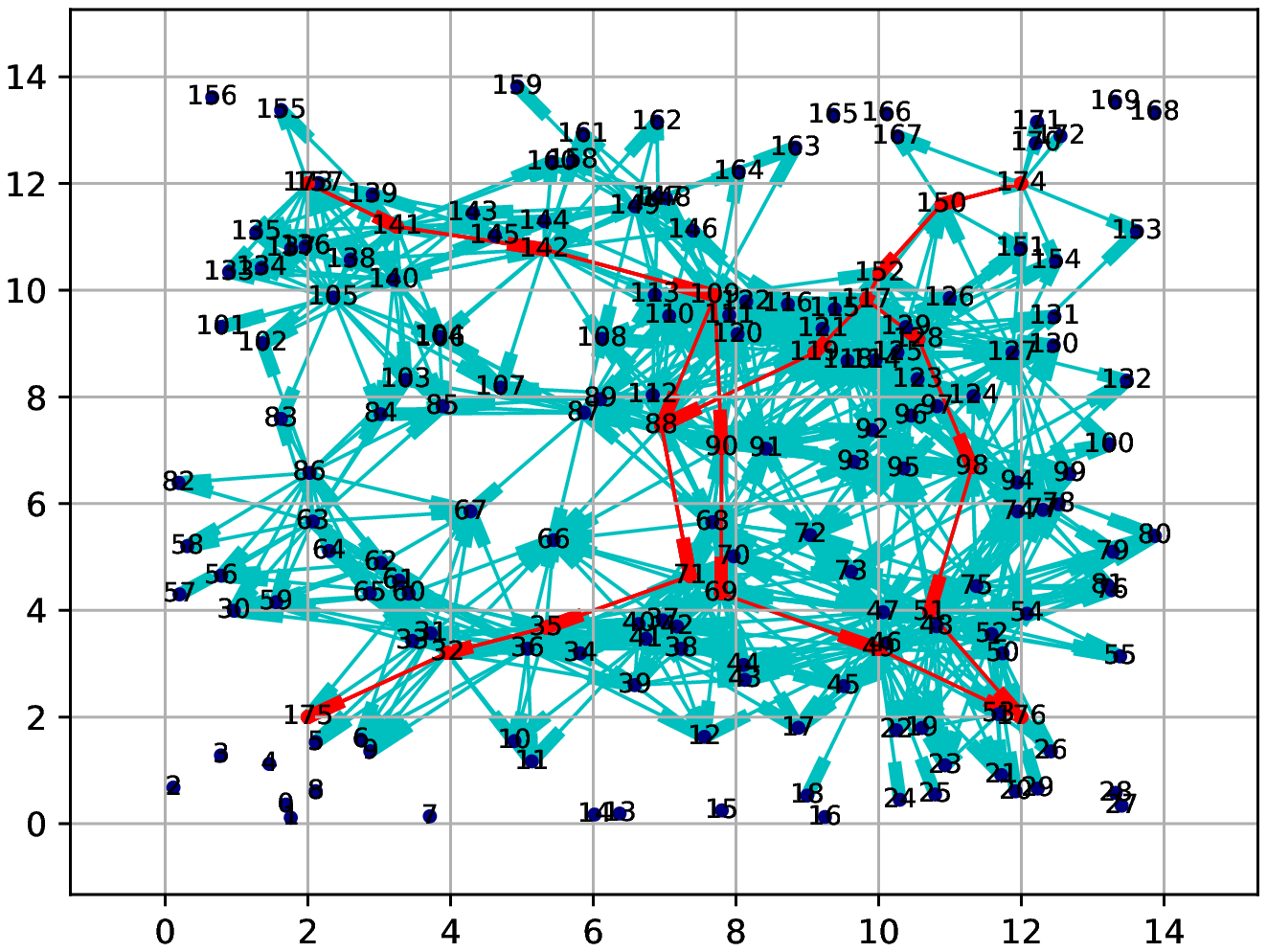}
        \caption{Step 60}
        \label{fig:step60}
    \end{subfigure}\\
      \begin{subfigure}[b]{0.48\textwidth}
        \includegraphics[width=\textwidth]{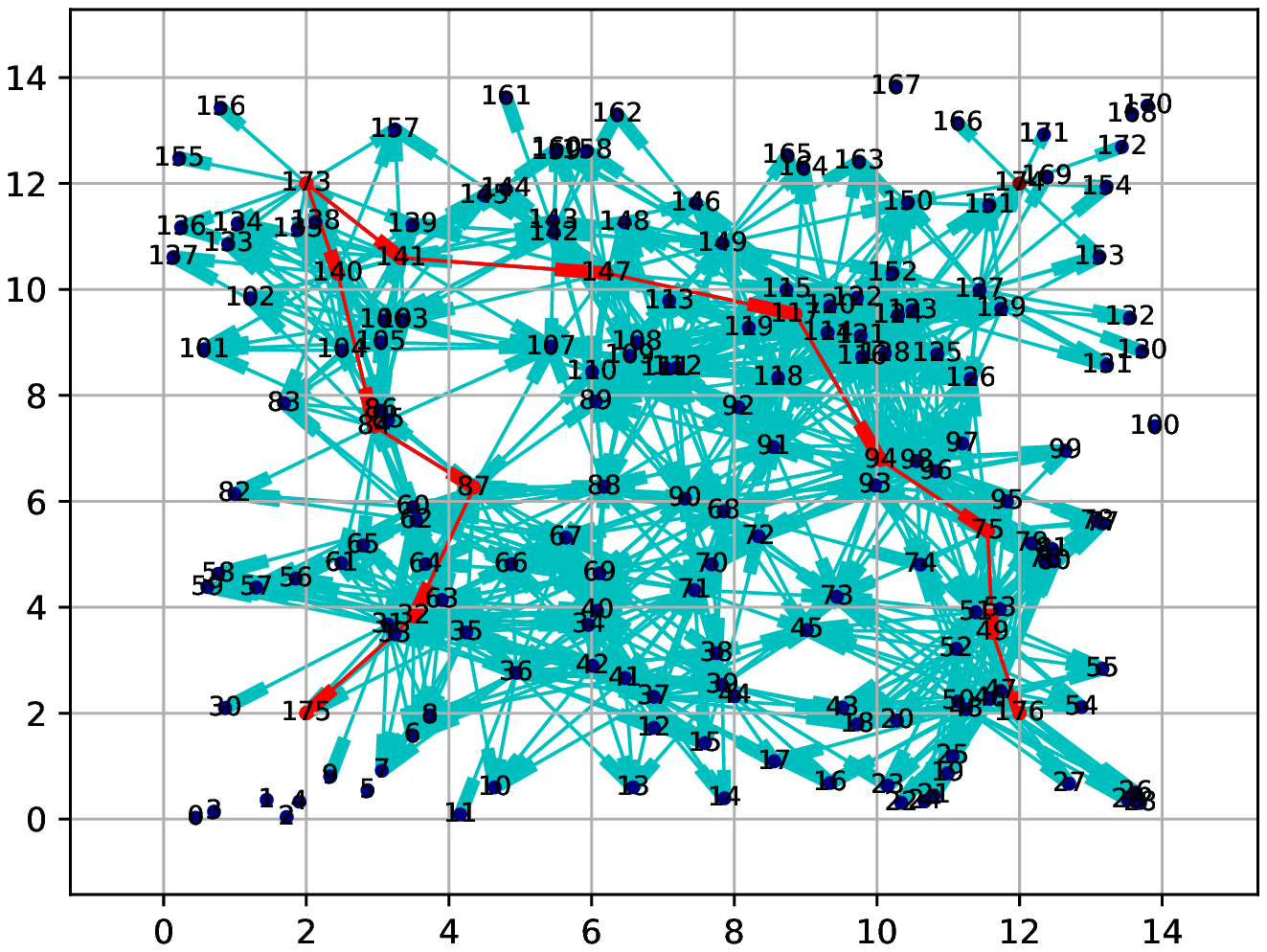}
        \caption{Step 120}
        \label{fig:step120}
    \end{subfigure}
        ~ 
    \begin{subfigure}[b]{0.48\textwidth}
        \includegraphics[width=\textwidth]{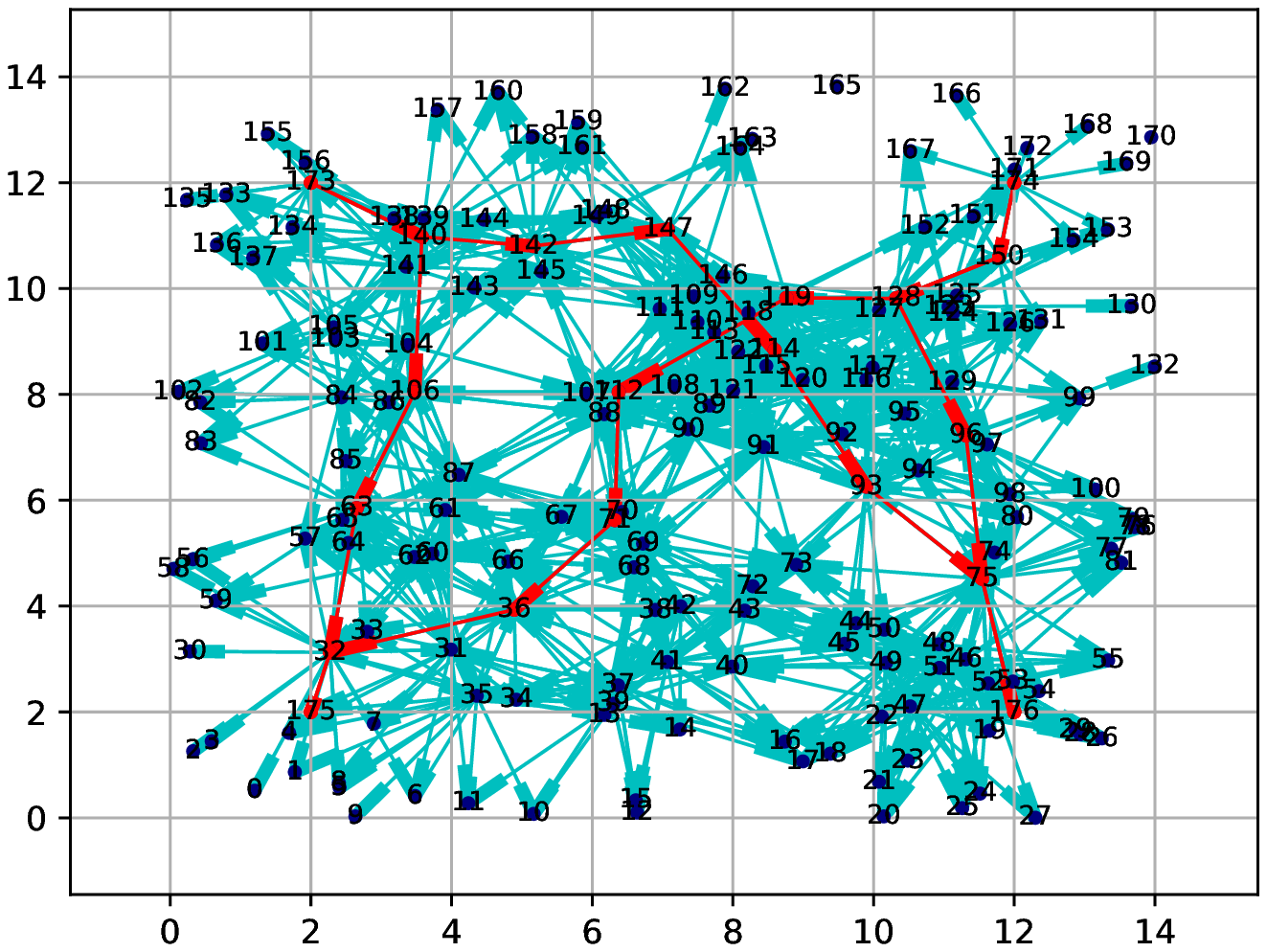}
        \caption{Step 150 (Final Network)}
        \label{fig:step150}
    \end{subfigure}
    \caption{
   Snapshots of the resulting network  across the  learning stage: the red  lines are the shortest paths  between sources and terminals   }
    \label{fig:nwsnapshot}
\end{figure*}
The learning process begins
at time step $0$
by initializing 
the transmission range of all nodes as zero. 
Then,  
nodes gradually increase their transmission ranges to make stable
connections from sources to terminals in 
the early stage of the learning process (e.g., between $0$ and $80$ time steps).
This is because 
the reward function defined in
\eqref{eqn:reward} is designed such that 
the throughput 
enhancement (i.e., $\phi_{\tau-1} - \phi_{\tau-2}$) is high enough
at this stage 
to take 
the action $a_{i,\tau-1}$ that 
enlarges 
the transmission range.
The corresponding changes in the radius of the transmission range 
between $0$ and $60$ steps are shown in  
Figure~\ref{fig:color}, and the actual connections are depicted in  
Figure~\ref{fig:step30} and Figure~\ref{fig:step60}.

As more learning stages are processed,  
the changes in the transmission range of each node  gradually slow down,  
and 
the transmission range of each node 
 eventually
converges to either zero or the maximum (i.e., $3.0$), as shown 
at time step
$150$.
This can be viewed as the \emph{activeness} of each node, i.e., 
an active node makes a connection with the maximum transmission range, and an inactive
node does not make a connection. 
Note that the node activeness is determined to improve the network throughput as can be seen  by comparing Figure~\ref{fig:step120}
and Figure~\ref{fig:step150}. Nodes  change their transmission range  to find  a shorter path from sources to terminals.
Thus, the proposed solution  determines which relay nodes  are essential for 
network throughput improvement and simultaneously minimizes power consumption by 
turning off potentially unnecessary  relay nodes. 



\subsection{Performance Comparison}

We next compare the performance of the proposed solution 
with  three
existing solutions. 

\begin{itemize}

\item
Value Iteration~\cite{kwon2017MDP}: A state-of-the-art model-based MDP solution with
the knowledge of  state transition probability and reward. This solution gives an
analytical solution to the 
near-optimal policy.
Then, the state transition matrix can be found using that policy, which allows the
system to analytically determine a stationary network.  In this experiment, the
optimality level of the policy is set to  $0.01$. 


\item
{TCLE}~\cite{xu2016}:
A state-of-the-art distributed 
solution for designing node connections. 
This solution enables a node to choose its transmission power by considering
the target
algebraic connectivity~\cite{gross2004} against transmission energy dissipation. 
In this experiment, the target algebraic connectivity  is set to $0.1$. 

\item
{Random Selection}: This solution enables a node to randomly select its 
action from  a given
action set. In this experiment, random selection used  the setting of 
$\epsilon = 1$.  

\end{itemize}

\begin{table*}[tb]
\caption{Performance comparison  }
\label{table:comp_small}
\vskip 0.15in
\begin{center}
\begin{small}
\begin{sc}
\begin{tabular}{c|ccccc}
\toprule
 & Proposed  & Value Iteration & Random & TCLE  \\
 &  & 
  & Selection & 
  \\
\midrule
\multicolumn{5}{c}{ Small Network:  $6.4 \times 10^3 [m^2]$} \\
\midrule
System Goodput [$Mbps$]&	461.08 & 427.15	&328.51&240.79\\
Connectivity Ratio&0.63	&	0.66&0.51&0.32\\
\midrule
\multicolumn{5}{c}{Large Network: $1.0 \times 10^4 [m^2]$ } \\
\midrule
System Goodput [$Mbps$]&	323.05 & 268.00 &202.93	&191.83\\
Connectivity Ratio&0.64	&	0.67&	0.47 & 0.38&\\
\bottomrule
\end{tabular}
\end{sc}
\end{small}
\end{center}
\vskip -0.1in
\end{table*}

Table~\ref{table:comp_small} shows the system goodput and connectivity ratio of the 
four distributed solutions with two  network sizes. 
At both sizes, the proposed solution achieves higher system goodput than value
iteration while maintaining a comparable connectivity ratio. 
Note that 
the value iteration method requires much more information 
than the proposed
solution; the state transition probability cannot be easily obtained in
real-world applications. Thus, the proposed solution 
is more practical than  value
iteration. 
The random selection method shows a lower performance than both the proposed solution and
value iteration because it does not include a learning process. 
The TCLE method shows the worst performance for both system goodput and connectivity ratio  because actions are taken without considering specific
source-to-terminal connections. TCLE considers only general network
connectivity in terms of algebraic connectivity.

\section{Conclusions}
\label{sec:conclusion}
In this paper, we have proposed a distributed solution 
based on the 
DDQN, which enables relay nodes to make decisions for multi-hop ad hoc network
formation with only partial observations 
in a network with many mobile relay nodes. 
The proposed solution can activate essential relay nodes and deactivate unnecessary
relay nodes,  leading to an autonomous node
activation system that can successfully build a network in a distributed manner. 
A deep reinforcement learning algorithm is deployed 
for decision-making at the
relay nodes, which update their wireless transmission ranges by observing the number of
nodes  in their current transmission ranges.  
The reward function includes network throughput and transmission power
consumption so that each relay node can choose its  action by explicitly 
considering the trade-off 
between network performance and its own power consumption. 
The DDQN used in the decision-making process can efficiently
optimize the Q-function. 
Our experimental results confirm that the network built by the proposed system
outperforms  those built by 
existing state-of-the-art solutions in terms of system
goodput and connectivity ratio.

\bibliographystyle{icml2018}
\bibliography{manuscript_v5}

\end{document}